\documentclass[twocolumn,aps,prb,graphicx,showpacs]{revtex4}
\usepackage{amsmath}
\usepackage{amsfonts}
\usepackage{amssymb}
\usepackage{graphicx}
\begin{document}
\title{Addendum: Theory of low-temperature Hall effect in stripe-ordered cuprates}
\author{Jie Lin and A. J. Millis}
\affiliation{Department of Physics, Columbia University, 
538 West 120th Street, New York, NY 10027}

\begin{abstract}
We supplement a previously published calculation of the Hall effect in stripe ordered cuprates by presenting results for the physically relevant sign of the potential arising from charge ordering. The new results   are more consistent with data, underscoring the importance of charge ordering in cuprates.
\end{abstract}
\maketitle

In a previous paper,\cite{Lin08} we studied the low-temperature Hall effect in a two-dimensional anti-phase stripe-ordered system and compared our results to existing experimental data.\cite{Daou09,Nakamura92} The calculations were based on a model introduced by one of us and Norman \cite{Millis07} involving fermions moving in two dimensions with dispersion $\varepsilon_k=-2t_1(\cos k_x+\cos k_y)+4t_2 \cos k_x \cos k_y-2t_3(\cos2k_x+\cos2k_y)$ (we chose $t_1= 0.38$eV, $t_2=0.32t_1$ and $t_3 = 0.5t_2$) and subject to a spin dependent potential of strength $V_s$ and wavevector $\mathbf{Q}$ chosen to be $(\pi,3\pi/4)$ and a spin independent ``charge" potential of amplitude $V_c$ and wavevector $2\mathbf{Q}$.  In that paper we considered only $V_c>0$ (the sign of $V_s$ is irrelevant). Very recently Vojta, Hackl and Sachdev have pointed out that  $V_c<0$ is the relevant case for cuprates. \cite{Vojta09,Hackl09} In this note we present calculations of the spin and charge density which confirm that this is the case and extend our Hall effect calculations to this situation. We find, as was anticipated in the Fermi surface plots of Ref \onlinecite{Millis07} and demonstrated for the Nernst effect by Ref \onlinecite{Hackl09} that changing the sign of $V_c$   makes the Hall conductance much more robustly electron-like, increasing the agreement between theory and experiment. 

\begin{figure}[htbp]
\centering
\includegraphics[width=0.9\columnwidth]{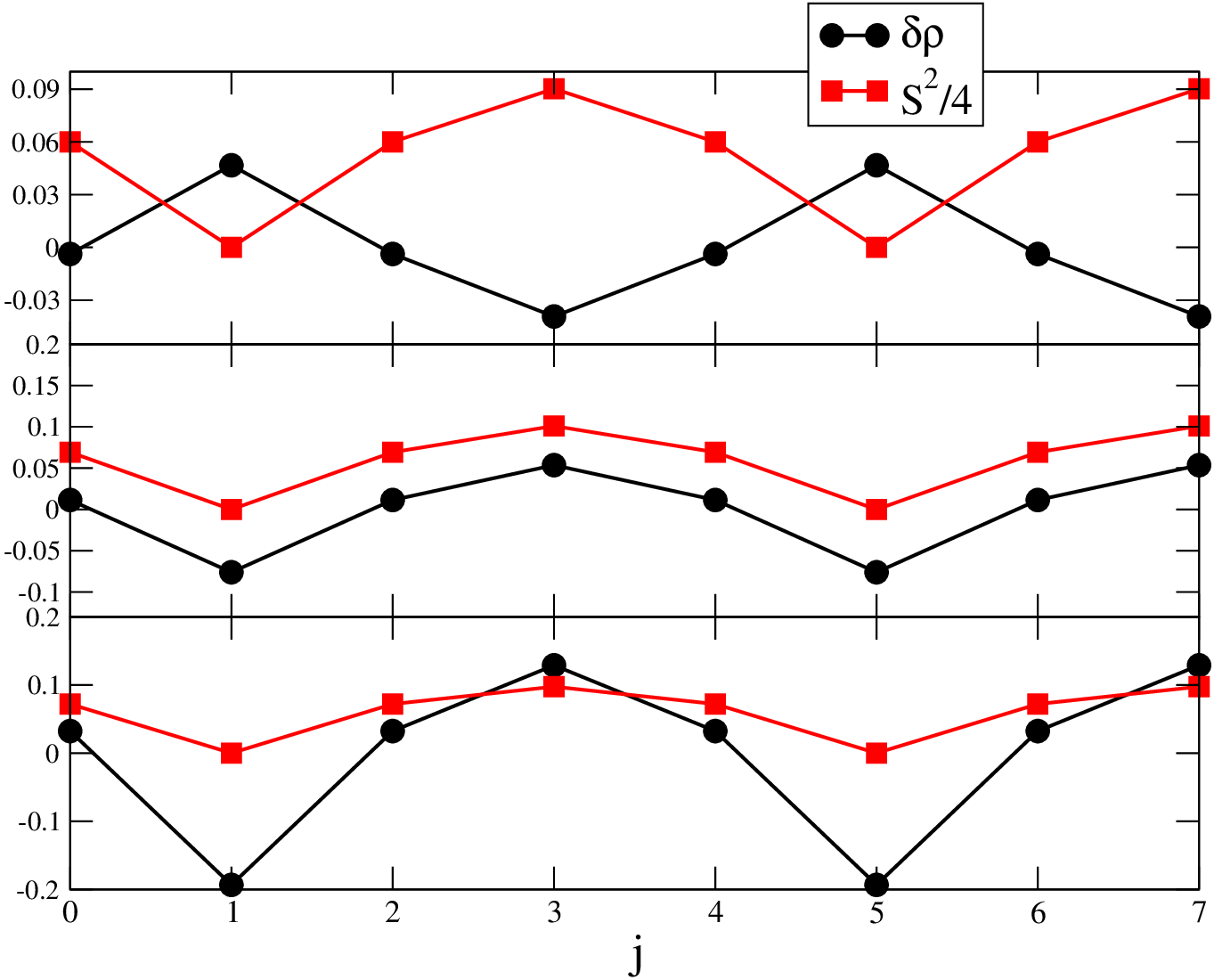}
\caption{\small{Real-space spin modulation multiplied by 1/4 (red squares) and charge modulation (black circles) along a row of lattice sites labeled by $j$ calculated for the 2-dimensional mean-field anti-phase stripe ordered state for  spin potential $V_s=0.25$eV and charge potential $V_c=0.08$eV (top), $0$ (middle) and $-0.08$eV (bottom) with doping fixed at $x\approx 1/8$.}}
\label{fig:density}
\end{figure}

We use the equations and methods of Ref \onlinecite{Lin08} which we do not repeat here. To compute the density we must evaluate the ground-state expectation value of the density operator ${\hat \rho}(j)$ for electrons  on site $j$ using the ground state wave function appropriate for electrons of spin $\sigma$. The periodicity we consider means that a single particle state at wavevector $\mathbf{k}$ in the irreducible Brillouin zone  is an $8$ component vector involving components $|\mathbf{k}+n\mathbf{Q}>$ with $n=0...7$.   ${\hat \rho}(j)$ is an $8\times8$ matrix with entries $<\mathbf{k}+n\mathbf{Q}|{\hat \rho}(j)|\mathbf{k}+m\mathbf{Q}>=\exp[i(m-n)\mathbf{Q}\cdot \mathbf{R}_j]$. The density of electrons of spin $\sigma$ on site $j$ is then calculated as  $\sum_{k,occ}<\psi_{k,\sigma}|{\hat \rho}(j)|\psi_{k,\sigma}>$ where the sum is over occupied states. 

In the presence of spin and charge backscattering the charge density and the square of the spin density oscillate with period $2Q$.  Fig \ref{fig:density} plots the  oscillating part of the charge density  $\delta\rho(j)=\rho_{\uparrow}(j)+\rho_{\downarrow}(j)-\rho_0$ and the square of the normalized spin density $S^2(j)=(\rho_{\uparrow}(j)-\rho_{\downarrow}(j))^2$  for positive, negative and vanishing $V_c$ with the chemical potential in each case adjusted to give a doping of $1/8$. The case $V_c=0$ already produces some charge modulation (putting the hole where the spin density is minimal); $V_c>0$ pushes the charge density modulation in the opposite direction (increasing the density on the row where the spin density vanishes) and $V_c<0$ makes even more holes on the site where the spin density vanishes. It is generally believed that in the high $T_c$ cuprates the hole density is maximal on the sites where the spin density vanishes; thus we conclude in agreement with Refs \onlinecite{Vojta09,Hackl09} that the $V_c<0$ case is the one which is relevant to the cuprates. Note that increasing the amplitude of $V_c$ to larger values would lead to unphysical $\delta\rho>1/8$, because this simple model does not include the Mott physics. 

\begin{figure}[htbp]
\centering
\includegraphics[width=1\columnwidth]{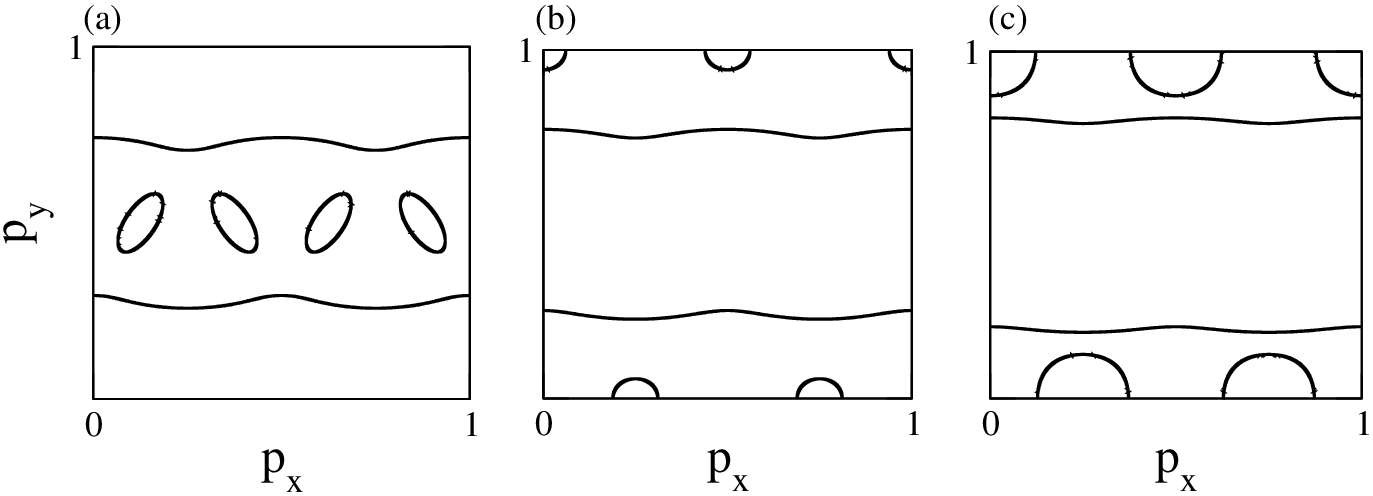}
\caption{\small{Fermi surface reconstruction in the stripe ordered state for $V=0.25$eV, and $V_c=0.1$eV (a), $V_c=0$ (b), and $V_c=-0.1$eV (c). The doping is fixed at $x\approx 0.125$, and the momentum is given in units of $\pi/a$.}}
\label{fig:fs}
\end{figure}

The backscattering potential produces a complicated Fermi surface, including electron and hole pockets and quasi one dimensional bands. Ref \onlinecite{Millis07} already showed that choosing a negative $V_c$ enhanced the electron pockets. We show in Fig \ref{fig:fs} the Fermi surfaces derived at $1/8$ doping for a fixed spin potential $V_s=0.25$eV and $V_c=0.1$ (left panel), $0$ (center panel) and $V_c=-0.1$eV (right panel). The stabilization of the electron pocket by the negative $V_c$ is evident. 

\begin{figure}[htbp]
\centering
\includegraphics[width=0.9\columnwidth]{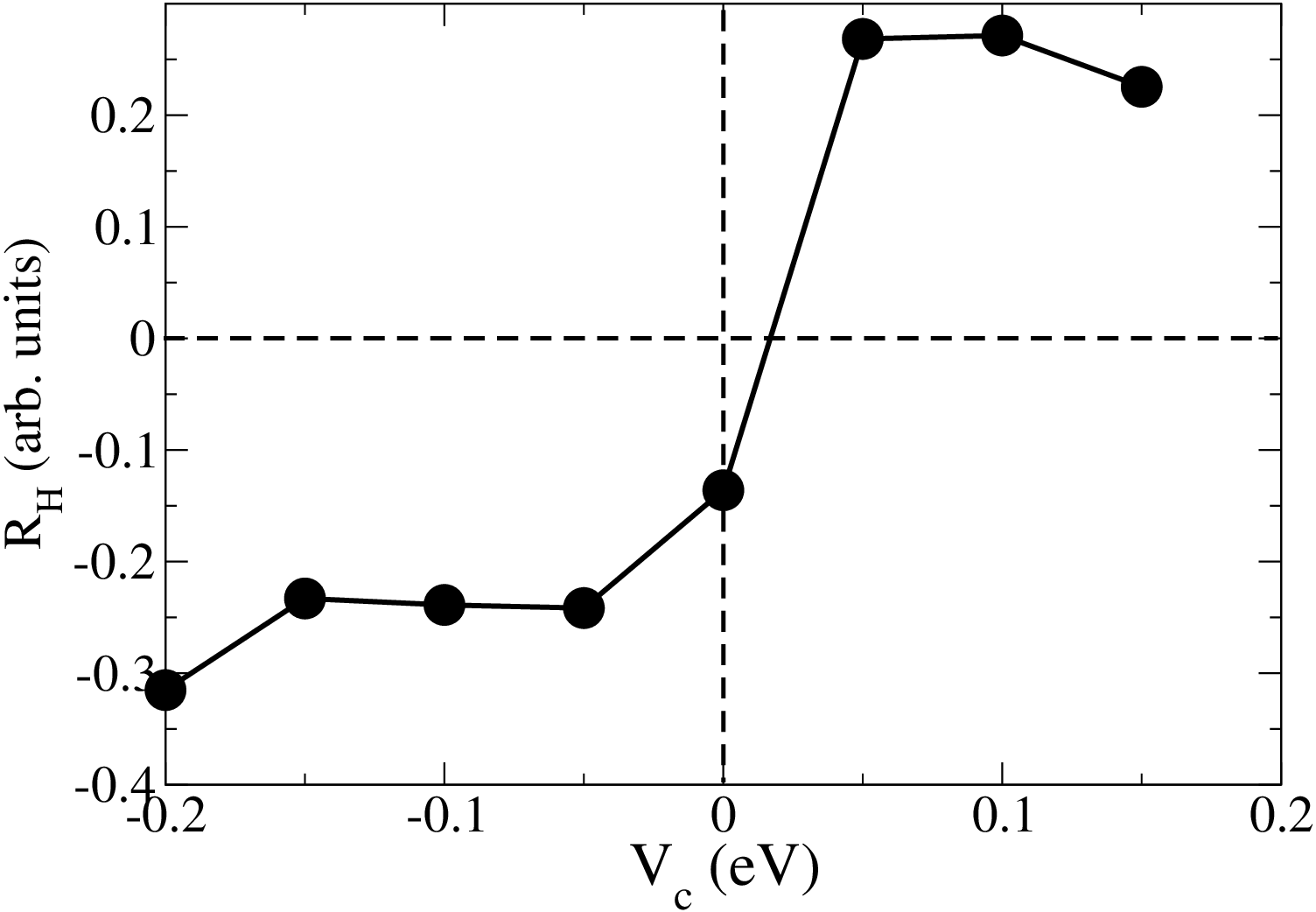}
\caption{\small{The Hall effect $R_H$ as a function of $V_c$ in the two-dimensional anti-phase stripe ordered state for $V_s=0.25$eV and doping $x=0.125$.}}
\label{fig:rh}
\end{figure}

Fig \ref{fig:rh} shows the dependence of the Hall coefficient on $V_c$ obtained using the methods of Ref \onlinecite{Lin08}. The dominance of the electron pocket in the calculations for $V_c<0$ leads to the electron-like sign of the Hall effect, while as $V_c$ is increased to positive values the sign rapidly turns hole-like. A characteristic feature of the experimental data is a robust electron-like signature in the Hall resistance which is favored by  the negative value of $V_c$ suggested by the charge density argument. The doubts expressed in Ref \onlinecite{Lin08} concerning the importance of the charge stripe potential are not warranted.

We thank M. Vojta for pointing out that our previous paper did not treat the relevant sign of $V_c$ and M. Norman for helpful discussions, and acknowledge support from NSF-DMR-0705847.

\end{document}